\documentclass[useAMS,usenatbib,referee]{biom}
\def\bSig\mathbf{\Sigma}

\usepackage{ulem}
\usepackage[utf8]{inputenc}
\usepackage{amsmath, amsfonts}
\usepackage{graphics, graphicx}
\usepackage{xcolor}
\usepackage{hyperref}
\usepackage{multirow}
\usepackage[caption=false]{subfig}

\allowdisplaybreaks

\title[Bayesian network-guided sparse regression with flexible varying effects]{Bayesian network-guided sparse regression with flexible varying effects}

\author{Yangfan Ren$^{1}$, 
Christine B.\ Peterson$^{2}$, and Marina Vannucci$^{1,*}$\email{marina@rice.edu} \\
$^{1}$Department of Statistics, Rice University, Houston, TX \\
$^{2}$Department of Biostatistics, The University of Texas MD Anderson Cancer Center, Houston, TX}

\begin{document}

\date{{\it Received January} 2024. {\it Revised 00} 0000.  {\it
Accepted 00} 0000.}

\pagerange{\pageref{firstpage}--\pageref{lastpage}} 
\volume{00}
\pubyear{2024}
\artmonth{January}

\doi{10.1111/j.1541-0420.2005.00454.x}

\label{firstpage}

\begin{abstract}
In this paper, we propose Varying Effects Regression with Graph Estimation (VERGE), a novel Bayesian method for feature selection in regression. Our model has key aspects that allow it to leverage the complex structure of data sets arising from genomics or imaging studies. We distinguish between the \textsl{predictors}, which are the features utilized in the outcome prediction model, and the subject-level \textsl{covariates}, which modulate the effects of the predictors on the outcome. We construct a varying coefficients modeling framework where we infer a network among the predictor variables and utilize this network information to encourage the selection of related predictors. We employ variable selection spike-and-slab priors that enable the selection of both network-linked predictor variables and covariates that modify the predictor effects. We demonstrate through simulation studies that our method outperforms existing alternative methods in terms of both feature selection and predictive accuracy. We illustrate VERGE with an application to characterizing the influence of gut microbiome features on obesity, where we identify a set of microbial taxa and their ecological dependence relations. We allow subject-level covariates including sex and dietary intake variables to modify the coefficients of the microbiome predictors, providing additional insight into the interplay between these factors.\\
\end{abstract}


\maketitle

\section{Introduction}
In this paper, we propose a novel Bayesian hierarchical regression model that enables the selection of both network-linked predictor variables and covariates that modify the predictor effects. Here, we distinguish between the \textsl{predictors}, which are the features utilized in the outcome prediction model, and the subject-level \textsl{covariates}, which modulate the effects of the predictors. 
Our method is motivated by applications to multivariate data sets arising from genomic and neuroimaging studies, where the observed predictors are linked by metabolic or functional networks. As an illustration of the utility of our method, we consider an application to a microbiome data set examining the interplay between the microbiome, obesity, and subject-level covariates \citep{wu2011linking}. In this context, the dependence network among the predictor variables describes ecological relationships between microorganisms inhabiting the same niche \citep{Kurtz2015}. Subject-level variables, including sex and dietary intake variables,  modify the influence of the microbiome features by regulating their activity and their effects on the host \citep{Leeming2021}. This effect may be partially driven by the production of dietary metabolites \citep{Sonnenburg2016}; however, many aspects of these relations are not completely understood. 

Our proposed model builds upon the framework of varying coefficient models \citep{Cleveland1991, Hastie1993}. Varying-coefficient models relax the assumption of linear effects in classical regression by allowing predictor effects to depend on factors that may modify their effects. This class of model has been extended to allow for high-dimensional covariates using spline- and tree-based approaches \citep{marx2009, burgin2015tree}. Additionally, Bayesian varying-coefficient models have been developed for scenarios with spatial or temporal dependence \citep{reich2010bayesian, scheipl2012spike}. These approaches generally focus on selecting either the main predictors or the modifying covariates. The simultaneous selection of predictors and covariates was first introduced in the pliable lasso \citep{tibshirani2020pliable} and extended by \cite{kim2021svreg} to account for grouping structure among the predictors via a weighted hierarchical penalty. In the Bayesian framework, \cite{ni2019bayesian} proposed a varying-sparsity regression model, which allows for subject-specific predictor selection and coefficient values. However, none of these approaches utilize information on the network among the predictor variables.

In our formulation of the varying coefficient modeling framework, we enable flexibility in the predictor effects by utilizing a Gaussian process prior, which allows the model coefficients to vary smoothly as a function of the observed covariates.  Furthermore, to achieve model sparsity, we rely on spike-and-slab priors for the selection of both the predictor and covariate effects. Our prior formulation allows to infer a network among the predictors and to utilize this information to encourage the selection of network-linked predictors. In order to infer the predictors network, we adopt the prior formulation proposed by \cite{wang2015scaling}, which imposes a mixture of normals on the off-diagonal elements of the precision matrix, along with an efficient blocked Gibbs sampling scheme. 
This approach directly represents edge selection, with respect to alternative shrinkage priors, such as the Bayesian graphical lasso  \citep{wang2012bayesian} or the graphical horseshoe prior \citep{li2019graphical}. For posterior inference, we design a stochastic search MCMC that requires careful consideration of how to handle the changing dimensions of the parameter space, cleverly dealing with the multiple layers of selection while ensuring good mixing. We also look into prediction. We call our proposed method VERGE (Varying Effects Regression with Graph Estimation).

We demonstrate through simulation studies that our method outperforms existing alternative methods in terms of both feature selection and predictive accuracy. We illustrate VERGE with an application to characterizing the influence of gut microbiome features on obesity, where we identify a set of microbial taxa and their ecological dependence relations. 
Our findings highlight bacterial genera with both protective and detrimental effects, and provide insight into how these effects are modulated by dietary intake and biological sex.

Section \ref{sec:methods} details our proposed VERGE approach and Section \ref{sec:posterior} discusses posterior inference and prediction. In Section \ref{sec:simulation}, we present simulation studies and comparisons. Section \ref{sec:case study} contains an application to microbiome data and Section \ref{sec:conclusion} conclusions and discussion.

\section{Methods}\label{sec:methods}

\subsection{Varying-effects regression model}\label{model}

Let $Y_i$ denote the observed response variable, and $\boldsymbol{X}_i = (X_{i1}, \ldots, X_{iP})$ denote the $P$-dimensional vector of predictors for subject $i=1,\ldots,n$. {We assume a joint distribution for the random variables $(Y, \boldsymbol{X})$. As in \cite{peterson2016joint}, our joint distribution $(Y, \boldsymbol{X})$ can be factorized as $(Y, \boldsymbol{X}) = f(Y|\boldsymbol{X})\cdot f(\boldsymbol{X})$, where $f(Y|\boldsymbol{X})$ is a regression model and $f(\boldsymbol{X})$ is a multivariate normal distribution.} Both the response variable $Y_i$ and predictors $\boldsymbol{X}_i$ are assumed to be centered. 
Additionally, for each subject $i$, we also observe a $K$-dimensional covariate vector $\boldsymbol{Z}_i = (Z_{i1}, \ldots,Z_{iK})$. Our proposed model allows for the effects of the predictors on the outcome to depend on specific covariates via a varying-effects regression model formulation where the coefficients of $X_{ij}$ change based on the value of $\boldsymbol{Z}_i$ as
\begin{equation}
y_i = \sum_{j=1}^pX_{ij}\boldsymbol{{\mu}}_j(\boldsymbol{Z}_i) + \varepsilon_i, \quad\varepsilon_i\sim \mathcal{N}(0,\tau^2),
\label{regression1}
\end{equation}
with {$\boldsymbol{\mu}_j(\boldsymbol{Z}_i)$} an unknown function of $\boldsymbol{Z}_i$, and $\varepsilon_i$'s iid white noise with variance parameter, $\tau^2$, on which we assume a standard conjugate inverse gamma prior as $\tau^2\sim\text{IG}(a_0, b_0)$. {Equation \eqref{regression1} represents a full model with no predictor selection; in the next subsection, we introduce our prior formulation that enables model sparsity.}

\subsection{Priors for variable and graph selection}\label{prior_pred}
Our model builds on {the Bayesian variable selection approach originally proposed by \cite{kuo1998variable}}.  We innovate on this framework in two key regards: we utilize network relations to link the probability of predictor selection and  allow the non-zero coefficients to vary as a smooth function of the covariate values. To represent the predictor selection, we introduce a set of latent indicator variables, $\boldsymbol{\gamma} = (\gamma_1, \ldots, \gamma_P)$ {, and write $\mu_j(\boldsymbol{Z}_i) = \gamma_j\boldsymbol{\beta}_j(\boldsymbol{Z}_i)$, assuming a priori independence between the indicators $\gamma_j$ and the effects $\boldsymbol{\beta}_j(\boldsymbol{Z}_i)$.} If $\gamma_j=0$, then $X_j$ has no effect on the response, as in the discrete spike-and-slab prior formulation \citep{vannucci2021discrete}. 
Thus, the varying-effects regression model in \eqref{regression1} can be written as
\begin{equation}
y_i = \sum_{j=1}^pX_{ij}\gamma_j\boldsymbol{\beta}_j(\boldsymbol{Z}_i) + \varepsilon_i, \quad\varepsilon_i\sim \mathcal{N}(0,\tau^2).
\label{regression2}
\end{equation}
{In this model formulation, $\gamma_j$ acts as an indicator for the relevance of the predictor $X_j$. When $\gamma_j = 1$, it implies that the corresponding $\beta_j(\boldsymbol{z})$ is included in the model; otherwise $\beta_j(\boldsymbol{z})$ is effectively zero. Although $\gamma_j$ and $\beta_j(\boldsymbol{z})$ are a priori independent, the MCMC sampling process captures their relationship, with $\gamma_j$ more likely to be 1 when the GP realization of $\beta_j(\boldsymbol{z})$ is significantly different from zero. Empirically, this model form is preferable to alternative spike-and-slab prior formulations \citep{george1997approaches} because it allows us to integrate out the coefficients $\beta_j(\boldsymbol{z})$, making the sampling process more efficient.}

Given our focus on both selecting a subset of explanatory predictors and understanding the interconnections among  these variables, we rely on the Gaussian graphical model to infer a network among the predictors and assume that $\boldsymbol{X}_i$ follow a multivariate normal distribution 
\begin{equation}
\boldsymbol{X}_i\sim \mathcal{N}(\boldsymbol{0}, \boldsymbol{\Omega}^{-1}),
\label{dist_x}
\end{equation}
where $\boldsymbol{0}$ is a $P$-vector of 0s, and $\boldsymbol{\Omega}$ is the precision matrix, which can be used to represent the conditional dependencies among predictors. Non-zero off-diagonal entries $\omega_{ij}$ in $\boldsymbol{\Omega}$ correspond to conditional dependence relations between the corresponding predictors, while $\omega_{ij} = 0$ indicates that predictors $i$ and $j$ are conditionally independent given the remaining variables. To identify a sparse set of dependence relations, we place a mixture prior on the entries in $\boldsymbol{\Omega}$ as proposed by \cite{wang2015scaling}:
\begin{equation}
    p(\boldsymbol{\Omega}\mid \theta) = \{C(\theta)\}^{-1}\prod_{i<j}\biggl\{(1-\pi)\mathcal{N}(\omega_{ij}\mid 0, \nu_0^2) + \pi \mathcal{N}(\omega_{ij}\mid 0, \nu_1^2)\biggl\}\prod_i \text{Exp}\big(\omega_{ii}\mid\frac{\lambda}{2}\big)I_{(\boldsymbol{\Omega}\in M^+)},
\label{graph_prior1}
\end{equation}
where $\theta = \{\nu_0,\nu_1,\lambda,\pi\}$ represents the set of prior hyperparameters, 
$C(\theta)$ is a normalizing constant,  and $\pi$ indicates the prior probability of edge selection. The prior distribution on the off-diagonal elements of $\boldsymbol{\Omega}$ is a mixture of normals, with $\nu_0$ and $\nu_1$ being set small and large, respectively. This allows a clear separation between selected edges, with values significantly different from zero, and non-selected edges, where  $\omega_{ij}$ is close to zero. The diagonal elements follow an exponential distribution with parameter $\frac{\lambda}{2}$. The final term $I_{(\boldsymbol{\Omega}\in M^+)}$ expresses the constraint that $\boldsymbol{\Omega}$ belongs to the cone of symmetric positive definite matrices.

The graph can also be represented using a set of binary latent variables $G = \{g_{ij}\}_{i<j} \in \{0,1\}$, where $g_{ij}=1$ indicates the presence of edge $(i,j)$ in the graph $G$. The model in \eqref{graph_prior1} can then be expressed as the hierarchical model
\begin{align}
    p(\boldsymbol{\Omega}\mid G,\nu_0,\nu_1,\lambda) &= \{C(G,\nu_0,\nu_1,\lambda)\}^{-1}\prod_{i<j}\mathcal{N}(\omega_{ij}\mid 0, \nu_{g_{ij}}^2)\prod_i \text{Exp}\big(\omega_{ii}\mid\frac{\lambda}{2}\big), \notag\\
    p(G\mid\theta) &= \{C(G,\theta\}^{-1}C(G,\nu_0,\nu_1,\lambda)\prod_{i<j}\{\pi^{g_{ij}}(1-\pi)^{1-g_{ij}}\}.
\label{graph_prior2}
\end{align}
For standardized data,
\cite{wang2015scaling} recommends setting $\pi=\frac{2}{p-1}$ and $\lambda = 1$, noting that edge selection tends to be insensitive to the choice of $\lambda$. For $\nu_0$ and $\nu_1$, \cite{wang2015scaling} observes that stable MCMC convergence is achieved with $\nu_0 \geq 0.01$ and $\nu_1 \leq 10$. Additional sensitivity analysis results are provided in \cite{wang2015scaling}.

Instead of using the conventional approach of employing an independent Bernoulli prior for the variable selection indicator $\boldsymbol{\gamma}$, we adopt the model proposed by \cite{peterson2016joint}, which utilizes a Markov random field (MRF) prior to link the selection of predictors according to their relations in the graph $G$ as
\begin{equation}
    p(\boldsymbol{\gamma}\mid G)\propto \exp(a\boldsymbol{1}'\boldsymbol{\gamma} + b\boldsymbol{\gamma}'G\boldsymbol{\gamma}),
\label{MRF}
\end{equation}
where $a$ and $b$ are scalar hyperparameters. This prior connects the variable inclusion to the inference of the dependence network, encouraging the selection of predictors that are connected with other relevant predictors. The parameter $a<0$ controls the prior probability of selecting a variable without accounting for information in the graph, while the parameter $b$ controls the extent to which a variable's inclusion probability is influenced by the inclusion of connected variables in the graph. As discussed in \cite{li2010bayesian}, $b$ should be carefully selected, as high values result in very dense models, a phenomenon known as phase transition.

\subsection{Priors on predictor effects and covariate selections}\label{prior_cov}

To allow for covariates to modulate the strength of the predictor effects, we employ the Gaussian process prior framework proposed by \cite{savitsky2011variable}. Specifically, for each predictor, the prior distribution of $\boldsymbol{\beta}_j(\boldsymbol{Z})$ is defined by a Gaussian process regression model
\begin{equation}
    \boldsymbol{\beta}_j(\boldsymbol{Z}) = f_j(\boldsymbol{Z}) + \boldsymbol{\delta}_j,
\end{equation}
where $\boldsymbol{Z}$ is an $n \times K$ matrix, and $f_j(\boldsymbol{Z})$ is a realization of a Gaussian process $f(\boldsymbol{Z})\sim\mathcal{N}(0,\boldsymbol{C}_j)$. The ``jitter" term $\boldsymbol{\delta}_j$ is distributed as $\mathcal{N}(0, \frac{1}{r_j}\boldsymbol{I}_n)$, and $r_j$ is a precision parameter with prior $\text{Ga}(a_r, b_r)$. We can integrate out $f_j(\boldsymbol{Z})$ to obtain the marginalized likelihood
\begin{equation}
\label{cov}
\boldsymbol{\beta}_j(\boldsymbol{Z})\sim GP(\boldsymbol{0}, \boldsymbol{C}_j + \frac{1}{r_j}\boldsymbol{I}_n).
\end{equation} 
As noted in \cite{neal1998}, the ``jitter" term is added to the covariance matrix to maintain the positive definite condition in computation. Given \eqref{cov}, by selecting an appropriate covariance matrix, we establish a non-linear relationship between the predictor effects $\boldsymbol{\beta}_j$ and the covariates $\boldsymbol{Z}$. Different covariance matrices can capture this relationship, as discussed in \cite{Rasmussen2006}. We adopt the single-term exponential covariance structure from \cite{savitsky2011variable}, for its simplicity and flexibility, which accommodates a wide range of linear and non-linear relationships. However, VERGE is general and is capable of using any valid kernel function, e.g. the Matern kernel. The covariance matrix $\boldsymbol{C}_j$ in our model comprises a constant term and an exponential term,
\begin{equation} \label{covariance}
\boldsymbol{C}_j = \frac{1}{\lambda_{aj}}\boldsymbol{J}_n + \frac{1}{\lambda_{zj}}\text{exp}(-\boldsymbol{M}),
\end{equation}
where $\boldsymbol{J}_n$ is an $n\times n$ matrix of 1’s and $\boldsymbol{M}$ is a matrix with entries $m_{ii'}$. Here, $m_{ii'}$ is defined as $(\boldsymbol{Z}_i - \boldsymbol{Z}_{i'})'\boldsymbol{P}(\boldsymbol{Z}_i - \boldsymbol{Z}_{i'})$, where $\boldsymbol{P}$ is the diagonal matrix  $\text{diag}(-\text{log} (\rho_{j1}, \ldots, \rho_{jK}))$, and $\rho_{jk}\in[0,1]$ is the parameter associated with covariate $\boldsymbol{Z}_{k}$ for $k = 1,\ldots,K$. 

To identify which specific covariates are important in modulating the effect of each predictor, we place spike-and-slab priors on the covariance parameters
\begin{equation}
p(\rho_{jk}\mid\Tilde{\gamma}_{jk})=\Tilde{\gamma}_{jk}\mathbf{I}[0<\rho_{jk}< 1] + (1-\Tilde{\gamma}_{jk})\delta_1(\rho_{jk}),
\end{equation}
for $j=1,\ldots,P$ and $k = 1,\ldots,K$, where $\delta_1$ represents a point mass distribution at 1, which translates to 0 once the log transformation is applied. The indicator variable $\Tilde{\gamma}_{jk}$ follows a Bernoulli distribution $\text{Ber}(\alpha_{jk})$. When $\Tilde{\gamma}_{jk} =1$, the magnitude of $\rho_{jk}\in(0,1)$ controls the smoothness of the function, while $\Tilde{\gamma}_{jk} =0$ indicates the covariate $\boldsymbol{Z}_k$ having no effect on the $j$th predictor, with $\rho_{jk} = 1$. Finally, we complete our model by assuming Gamma priors on the scaling parameters of equation \eqref{covariance} as $\lambda_{aj}\sim\text{Ga}(a_\lambda, b_\lambda)$ and $\lambda_{zj}\sim\text{Ga}(a_z, b_z)$ for $j = 1,\ldots,P$. Due to the sensitivity to scaling of prior \eqref{covariance}, normalizing the covariates $\boldsymbol{Z}$ to the unit cube is recommended by \cite{savitsky2011variable}.

\section{Posterior inference}
\label{sec:posterior}
Given that the posterior is intractable, we utilize Markov chain Monte Carlo (MCMC) methods to sample parameters from the posterior distribution. To avoid directly sampling the realizations for $\boldsymbol{\beta}_j(\boldsymbol{Z})$, and to minimize uncertainty, we integrate out $\boldsymbol{\beta}_j(\boldsymbol{Z})$ for $j = 1,\ldots,P$. Since we have spike-and-slab priors on both the predictor coefficients and the covariance parameters for each covariate, our MCMC scheme requires careful consideration of how to handle the changing dimensions of the parameter space. To deal with the multiple layers of selection and ensure good mixing, we incorporate both between-model moves (where we update the predictor or covariate selection) and within-model moves (where we update the parameters while keeping the predictor and covariate selection fixed). The details of the MCMC scheme are provided in Web Appendix A.

\subsection{Variable and edge selections}
For variable selection, as recommended by \cite{barbieri2004optimal}, we utilize the median probability model, which includes variables with a marginal posterior probability of inclusion (PPI) of at least 0.5. The marginal PPI for each predictor $j$ is determined by calculating the frequency of inclusion in the model across the post burn-in MCMC samples, represented as $\delta_j = \frac{\sum_{t=1}^N\textbf{I}(\gamma^{(t)}_j=1)}{N}$, where $N$ is the total number of samples. For each covariate $Z_{jk}$, the PPI is calculated as the number of iterations where $\boldsymbol{Z}_{jk}$ is included in the model out of the total number of iterations where its corresponding $\boldsymbol{X}_j$ was selected. As an alternative, the expected false discovery rate (FDR) can be used to set a threshold for the PPIs. This is calculated as $\text{FDR}(\kappa) = \frac{\sum_{j=1}^P\textbf{I}(\delta_j>\kappa)(1-\delta_j)}{\sum_{j=1}^P(\delta_j>\kappa)},$ where $\kappa \in (0, 1)$ is the threshold value.  Furthermore, for edge selection, following the approach from \cite{wang2015scaling} and \cite{peterson2016joint}, we adopt the posterior median graph for the selection of the graph structure. Specifically, we select edges that have a marginal PPI greater than 0.5. 

\subsection{Prediction}
To perform prediction, we follow the method in \cite{Rasmussen2006} to incorporate the information provided by the training data about the function. Let $\boldsymbol{X}^*$ and $\boldsymbol{Z}^*$ be the predictors and covariates for future observations, and $\boldsymbol{\beta}^* = \boldsymbol{\beta}(\boldsymbol{Z}^*)$ represent the corresponding $n^*\times 1$ latent vector. We then consider the joint distribution 
\begin{equation*}
    \begin{bmatrix}
        \boldsymbol{\beta}\\
        \boldsymbol{\beta}^*
    \end{bmatrix}\sim\mathcal{N}
    \begin{pmatrix}
        \begin{bmatrix}
            \boldsymbol{0}\\
            \boldsymbol{0}
        \end{bmatrix},
        \begin{bmatrix}
            \boldsymbol{C}_{(\boldsymbol{Z},\boldsymbol{Z})} & \boldsymbol{C}_{(\boldsymbol{Z},\boldsymbol{Z^*})}\\
            \boldsymbol{C}_{(\boldsymbol{Z^*},\boldsymbol{Z})} & \boldsymbol{C}_{(\boldsymbol{Z^*},\boldsymbol{Z^*})}
        \end{bmatrix}
    \end{pmatrix},
\end{equation*}
where $\boldsymbol{C}_{(\boldsymbol{Z},\boldsymbol{Z^*})} := \boldsymbol{C}_{(\boldsymbol{Z},\boldsymbol{Z^*})}(\boldsymbol{\Theta})$ denotes the $n\times n^*$ covariance matrix calculated for all pairs of training and test points, {$\boldsymbol{\Theta} = \{\boldsymbol{\Theta}_1,\ldots,\boldsymbol{\Theta}_P\}$, and $\boldsymbol{\Theta}_j = \{\Tilde{\boldsymbol{\gamma}}_j, \boldsymbol{\rho}_j, \lambda_{aj}, \lambda_{zj}, r_j\}$ for $j = 1,\ldots,P$}. The expectation of the conditional joint predictive distribution for $\boldsymbol{\beta}^*\mid \boldsymbol{\beta}$ is $\boldsymbol{C}_{(\boldsymbol{Z}^*,\boldsymbol{Z})}\boldsymbol{C}_{(\boldsymbol{Z},\boldsymbol{Z})}^{-1}\boldsymbol{\beta}$, which we can estimate based on the MCMC samples as  
\begin{equation}
        \hat{\boldsymbol{\beta}}^*_j(\boldsymbol{\Theta}^{(t)}) := \boldsymbol{C}_{(\boldsymbol{Z}^*,\boldsymbol{Z})}(\boldsymbol{\Theta}^{(t)})\boldsymbol{C}_{(\boldsymbol{Z},\boldsymbol{Z})}^{-1}(\boldsymbol{\Theta}^{(t)})\hat{\boldsymbol{\beta}}_j,
    \label{cov_pred}
    \end{equation}
where $\hat{\boldsymbol{\beta}}_j = \frac{\sum_{t=1}^{N}\textbf{I}(\gamma^{(t)}_j = 1)\Tilde{\boldsymbol{\beta}}_j(\boldsymbol{Z})}{\sum_{t=1}^N\textbf{I}(\gamma^{(t)}_j = 1)}$, when the $j$th predictor is included in the model. Here, $\Tilde{\boldsymbol{\beta}}_j(\boldsymbol{Z})$ represents the sampled values from each iteration. We can then obtain the estimated response value $\hat{\boldsymbol{y}}^*=\frac{1}{L}\sum_{t=1}^L\bigl(\sum_{j=1}^P\mathbf{I}(\Bar{\gamma}_j>0.5)\boldsymbol{X}^*_j\hat{\boldsymbol{\beta}}^*_j(\boldsymbol{\Theta}^{(t)})\bigl)$, where $\Bar{\gamma}_j$ represents the marginal PPI for the $j$th predictor, and $L$ is the total number of samples where all predictors with marginal PPI greater than 0.5 are selected. This involves averaging over the MCMC samples to obtain the final estimate. Note that only covariates that have been selected based on the marginal PPIs are included in the computation of the covariance matrices in \eqref{cov_pred}. Moreover, as suggested by \cite{neal1998},  we rely on the Cholesky decomposition for computing $\boldsymbol{C}_{(\boldsymbol{Z},\boldsymbol{Z})}^{-1} $ in  \eqref{cov_pred}.

\section{Simulation study}
\label{sec:simulation}

\subsection{Simulation setup}
\label{sim_setup}
In our simulation design, we first construct the graph representing the dependence relations among the predictor variables. We create a sparse network with clusters of correlated features, similar to \cite{li2008network} and \cite{peterson2016joint}. Predictors are represented as clusters of genes, including a transcription factor and its regulated genes. Our graph consists of $P = 60$ nodes, divided into 12 clusters. Each cluster contains one primary node functioning as a hub, connected to four remaining nodes in the cluster, resulting in a network with 48 total edges.

The predictor variables $\boldsymbol{X}_i$ are sampled from a multivariate normal distribution with mean zero and covariance matrix $\mathbf{\Sigma}_G$. The matrix $\mathbf{\Sigma}_G$ has unit variances for each predictor. Within each cluster, we set a correlation of 0.7 between the primary node and the four subsidiary nodes, and the correlations among the subsidiary nodes are fixed to $0.7^2$. This results in a sparse graph structure with edges limited to within-cluster connections, where each primary node is connected to all four subsidiary nodes.

We assume 10\% of the predictors are relevant to the outcome, resulting in $P_{\text{true}} = P/10=6$. The response variable $y_i$ is generated using the linear model $y_i=\sum_{j=1}^6X_{ij}\boldsymbol{\beta}_j(\boldsymbol{Z}_i) + \varepsilon_i,$ for $i = 1,\ldots,n,$ where $\varepsilon_i\sim \mathcal{N}(0,1)$. We specify $n=200$ as the number of training samples for parameter estimation, and $n_t=50$ as the number of test samples for the evaluation of prediction performance. The covariates $\mathbf{Z}_k$ are randomly sampled from Unif$(-1,1)$ with $K=3$. For the predictor effects $\boldsymbol{\beta}_j(\boldsymbol{Z})$, we consider different generating functions, including constant, linear, and non-linear forms. The true values of the $\boldsymbol{\beta}_j(\boldsymbol{Z})$'s are defined as follows:
\begin{eqnarray}
\label{functions}
    \boldsymbol{\beta}_1(\boldsymbol{Z}_1) &=& 0.3,~
    \boldsymbol{\beta}_2(\boldsymbol{Z}_2) = 2\text{sin}(\pi \boldsymbol{Z}_2),~
    \boldsymbol{\beta}_3(\boldsymbol{Z}_3) = 2\boldsymbol{Z}_3^2-1,~
    \boldsymbol{\beta}_4(\boldsymbol{Z}_4) = -2\boldsymbol{Z}_4,\\
    \boldsymbol{\beta}_5(\boldsymbol{Z}_5) &=& 2\text{cos}(\pi \boldsymbol{Z}_5),~
    \boldsymbol{\beta}_6(\boldsymbol{Z}_6) = -2\mathcal{N}(\boldsymbol{Z}_6\mid 0.3, 0.3^2)-3\mathcal{N}(\boldsymbol{Z}_6\mid -0.5, 0.3^2),~
    \boldsymbol{\beta}_j(\boldsymbol{Z}_j) =0, \nonumber
\end{eqnarray}
for $j=7,\ldots, 60$, where $\boldsymbol{Z}_j$, which represents the covariate influencing the $j$th predictor, is randomly selected from the $K=3$ covariates. We exclude categorical functions in this setup since not all methods in our simulation studies are designed for handling categorical covariates. However, VERGE is effective with binary and categorical covariates, as shown in the application section. Finally, we center $\boldsymbol{y}$ and standardize the predictors $\boldsymbol{X}$ and covariates $\boldsymbol{Z}$ to ensure stable results when applying the graphical model \citep{wang2015scaling} and the Gaussian process model \citep{savitsky2011variable}. 

{Parameter settings and sensitivity analyses are discussed in Web Appendix B.}

\subsection{Comparative analysis}
\label{sim_comp}
{To characterize the impact of the components of our proposed VERGE model, including the incorporation of graph information and the selection of predictors, we consider two reduced forms of our model as comparators. In the first reduced model, we omit the second term in equation \eqref{MRF}, so that no graph information is incorporated under the prior. In this model, the prior on $\boldsymbol{\gamma}$ simplifies to an independent Bernoulli. In the second reduced model, we include a GP prior on $\beta_j(\boldsymbol{Z})$, but do not perform selection of the primary predictors.}

Furthermore, we compare VERGE with two established methods: the pliable lasso method \citep{kim2021svreg}, implemented in the R package \texttt{svreg}, and the spline-based Bayesian Hierarchical Varying-Sparsity Regression (BEHAVIOR) model \cite{ni2019bayesian}. The pliable lasso was fit with two penalty parameters selected using 5-fold cross-validation on the training data. The Bayesian models were run in Matlab Release 2022b. For the BEHAVIOR model, 150,000 iterations for burn-in and 150,000 for inference were needed to reach convergence, using the Matlab code provided by \cite{ni2019bayesian}. Our VERGE model required 60,000 iterations, with the first 30,000 for burn-in. {The acceptance rates for the Level 1 moves were 5\% for the “Add” proposal, 45\% for the “Delete” proposal, and 25\% for the “Keep” proposal. Despite the lower acceptance rate for the “Add” move, the rates for “Keep” and “Delete” moves facilitate updates in the GP kernel and improve mixing.} The acceptance rates for updating covariate inclusion in the within-model move ranged from 40-60\%. We evaluated the Pearson correlation for the PPIs of both predictor and covariate selections from two independent chains and found good indications of convergence.

We assess the performance of the models using the following metrics: the true positive rate (TPR), false positive rate (FPR), $\text{F}_1$ score, Matthews correlation coefficient (MCC), and area under the ROC curve (AUC), for both variable and covariate selection. Additionally, we employ the mean squared prediction error (PMSE) as a measure to evaluate the predictive performance of the models. Specifically, the metrics are defined as $\text{TPR} = \frac{\text{TP}}{\text{TP} + \text{FN}}$, $\text{FPR} = \frac{\text{FP}}{\text{FP} + \text{TN}}$, $\text{F}_1 = \frac{2\text{TP}}{2\text{TP} + \text{FP} + \text{FN}}$,  $\text{MCC} = \frac{\text{TP}\times\text{TN} - \text{FP}\times\text{FN}}{(\text{TP} + \text{FP})(\text{TP} + \text{FN})(\text{TN} + \text{FP})(\text{TN} + \text{FN})}$ and $\text{PMSE} = \frac{1}{n_t}\sum_{i=1}^{n_t}(\hat{y}_i - y_{\text{test},i})^2$, where TP, TN, FP, and FN denote the true positives, true negatives, false positives and false negatives, respectively and $n_t$ is the sample size for the test data. For the pliable lasso, which has two tuning parameters that regulate the penalties for predictor and covariate selections, the AUC calculation requires varying one of these parameters while selecting the second parameter through a five-fold cross-validation at each level of the first parameter. AUCs for VERGE and BEHAVIOR are calculated by changing the thresholds for the PPIs. 

\begin{table}
    \caption{Simulation results for predictor selection and prediction accuracy {for $n=200$, $P=60$, $K=3$}. Methods compared include the pliable lasso, BEHAVIOR, VERGE, {and two reduced versions of VERGE: one with no graph selection, and one with no predictor selection}. Performance metrics evaluated are the True Positive Rate (TPR), False Positive Rate (FPR), $\text{F}_1$ Score, Matthews Correlation Coefficient (MCC), the Area Under the ROC Curve (AUC) and Mean Squared Prediction Error (PMSE).}
    \begin{center}
    \begin{tabular}{lccccc}
      \Hline
     & pLasso & BEHAVIOR & VERGE & {No graph} & {No predictor selection} \\
     \hline
    TPR  & 0.460(.265) & 0.853(.055) & {0.960}(.073) & {0.833(.068)} & -  \\ 
    FPR  & 0.157(.190) & 0.002(.006) & {0.001}(.004) & {0.002(.005)}  & -  \\
    MCC  & 0.340(.194) & 0.905(.046) & {0.974}(.051) & {0.896(.056)}  & -  \\ 
    F1   & 0.356(.163) & 0.911(.042) & {0.975}(.047) & {0.902(.052)} & -  \\ 
    AUC  & 0.381(.159) & 0.966(.027) & {0.999}(.001) & {0.972(.033)} & - \\ 
    PMSE  & 6.657(2.195) & 1.326(.199) & {1.278}(.260) & {1.295(.276)} & -{7.125(2.220)} \\
    \Hline
    \end{tabular} 
    \end{center}
    \label{simulation_res_pred}
\end{table}

\begin{table}
    \caption{Simulation results for covariate selection for $n=200$, $P=60$, $K=3$. Methods compared include the pliable lasso, BEHAVIOR,
   VERGE, {and two reduced versions of VERGE: one with no graph selection, and one with no predictor selection}. Performance metrics evaluated are the True Positive Rate (TPR), False Positive Rate (FPR), $\text{F}_1$ Score, Matthews Correlation Coefficient (MCC), the Area Under the ROC Curve (AUC) and Mean Squared Prediction Error (PMSE). Note: some MCC values for the pliable lasso are omitted due to zero denominators.}
    \begin{center}
    \begin{tabular}{lccccc}
      \Hline
     & pLasso & BEHAVIOR & VERGE & {No graph}  & {No predictor selection} \\
     \hline
    TPR  &  0.280(.271) & {0.992}(.040) & {0.992}(.040) & {0.984(.055)} & {0.800(.062)} \\ 
    FPR  & 0.018(.050) & {0.001}(.001) & 0.007(.002) & {0.001(.001)} & {0.011(.012)} \\
    MCC  &  -           & {0.986}(.041) & 0.985(.035) & {0.988(.034)} & {0.758(.043)} \\ 
    F1   & 0.263(.241) & {0.985}(.042) & {0.985}(.036) & {0.988(.035)} & {0.764(.042)} \\ 
    AUC  & 0.075(.084) & 0.964(.018) & {0.996}(.022) & {0.992(.029)} & {0.995(.007)} \\ 
      \Hline\end{tabular} 
    \end{center}
    \label{simulation_res_cov}
\end{table}

We first considered a simulated setting with $n = 200, P = 60, K = 3$. Results {for predictor selection} are shown in Table \ref{simulation_res_pred} {and those for covariate selection are shown in Table \ref{simulation_res_cov}}, across 25 simulated data sets. {In comparison to the full VERGE model, the predictor selection for the reduced model with no graph information exhibited a lower TPR. Upon examining each predictor, this decrease is attributed to the extremely low TPR (0.080) for the predictor with a small constant effect. Without graph information, the reduced model fails to identify this effect. However, the covariate selection and PMSE remain similar. The reduced model variant with no selection of the primary predictors achieved acceptable performance in covariate selection, demonstrating its ability to identify most of the covariate effects. However, it led to a high PMSE, likely due to overfitting caused by the absence of predictor selection, as all predictors were considered in the model.}

With respect to competing methods, results indicate that VERGE performs well in both predictor and covariate selection, notably outperforming existing methods in predictor selection due to the integration of graph information. This helps in identifying connected predictors with slightly weaker effects. Additional simulations were conducted with more covariates ($K = 6$) and a smaller sample size ($n = 100$), with detailed results reported in Web Appendix C. Overall, VERGE achieves TPRs that are either 1 or cloase to 1 across all scenarios, with FPRs below 0.001. In terms of covariate selection, VERGE is comparable with BEHAVIOR. However, VERGE experiences a slight reduction in performance with an increase in covariates or a decrease in sample size, while BEHAVIOR shows a significant decrease in smaller sample scenarios but maintaining strong performance
with increased covariates. The pliable lasso, limited to identifying only linear covariate effects, exhibits lower accuracy across all scenarios. Additionally, its penalty parameters, optimized for prediction error through cross-validation, are not geared towards predictor selection, as discussed in \cite{meinshausen2006high}). In predictive performance, our method has the lowest PMSE in all scenarios, whereas the pliable lasso has the highest.

\begin{figure}
\centerline{
\includegraphics[width=18cm]{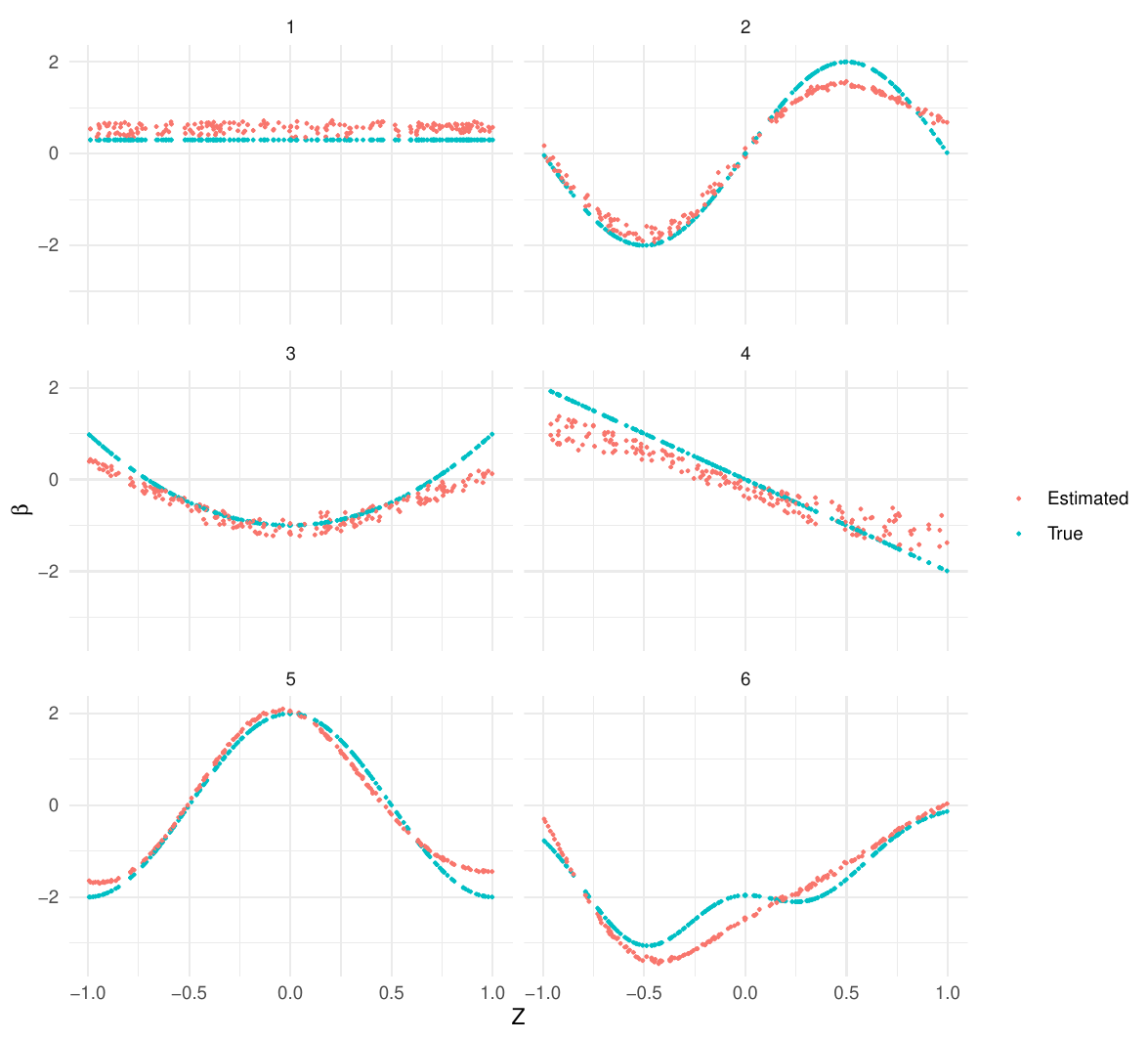}}
\caption{True and estimated coefficients for simulation study by generating functions.} 
\label{Sim_plot}
\end{figure}

Figure \ref{Sim_plot} displays true and estimated coefficients for each underlying function listed in equation \eqref{functions}, from one simulated dataset. The estimated coefficients were calculated by averaging the sampled $\boldsymbol{\beta}$'s from the third step of the MCMC algorithm, as detailed in Web Appendix A. VERGE effectively recovers all function types, including constant, linear, and non-linear, closely matching the true values. We also evaluated the average TPRs for predictor selection across all methods by generating function. Focusing on the base scenario ($n = 200$, $P = 60$, $K = 3$), results in Web Table 3 show that our method performs well across all functions. BEHAVIOR excels with covariate-dependent effects but struggles with small constant coefficients. The pliable lasso only performs well with functions that do not cross 0. VERGE's ability to detect small constant coefficients likely benefits from the graph structure guiding the predictor selection.

\section{Case study}
\label{sec:case study}
The gut microbiome, consisting of trillions of bacteria, plays a critical role in extracting energy from the diet and influences human health outcomes, including obesity \citep{Turnbaugh2006}. However, the mechanisms behind this link are not fully understood. In this case study, we consider how the effect of gut microorganisms on body mass index (BMI) may be modulated by covariates such as sex and dietary intake, using the COMBO dataset originally described by \cite{wu2011linking}. The data were obtained from a cross-sectional study of 98 healthy volunteers, where stool samples were analyzed using 16S rRNA gene segments via 454/Roche pyrosequencing. Additionally, diet and demographic information, including age, sex, and BMI, were collected. This dataset was previously analyzed by \cite{lin2014variable} and \cite{zhang2021bayesian}, who focused on microbiome feature selection without considering potential covariate effects.

In our analysis, we utilized the reprocessed data from \cite{zhang2021bayesian}, which used the updated SILVA rRNA database to assign the sequences to taxonomy. We obtained 156 genera from 1763 OTUs using the R package \texttt{phyloseq}, focusing on 69 taxa with an average abundance $\geq 0.1\%$. Given the compositional nature of the data, the features were transformed using a centered log-ratio transform \citep{Aitchison1982} prior to downstream analysis. Since we are interested in predicting obesity as a health outcome, we used BMI as our response variable. We selected sex, total fat (tfat), and total fiber measured using the AOAC method \citep[aofib, ][]{McCleary2010} as the candidate covariates that could modify the effects of the microbiome features on BMI. The response variable was centered, and both the predictors and covariates were normalized in the analysis.

We fit the model to the data using parameter settings similar to those in our simulation study, but with slight modifications. To account for the relatively weaker signal compared to simulated data, we increased the prior probability of predictor inclusion parameter $a$ to $\text{log}(0.22)$. We also chose smaller values for the hyperparameters $\nu_0$ and $\nu_1$ for graph estimation, setting them at 0.1 and 5, respectively. The MCMC simulations included a burn-in of 100,000 iterations, followed by 100,000 iterations for analysis. Similar to the simulation study, we selected predictors and edges with marginal PPIs above 0.5. For covariate selection, we used a cutoff of 0.5, rather than controlling FDR, due to the weaker signal in real data.

The VERGE model identified 11 genera influencing BMI, as shown in Table \ref{realdata_sel}. It lists the genera, their covariates, and average coefficients across 98 subjects. Four genera exhibited protective effects with a negative association with BMI, including two from the family \textsl{Lachnospiraceae} and two from the family \textsl{Ruminococcaceae}. Previous studies have noted the depletion of these families in obese adults, suggesting they digest dietary fiber into short chain fatty acids, potentially modulated by diet \citep{Peters2018, Vacca2020}. Specifically, both \textsl{Ruminococcaceae} genera showed dietary-dependent effects. Conversely, genera like \textsl{Catenibacterium} and \textsl{Megasphaera}, linked to increased BMI in prior research \citep{Pinart2021}, were also identified.

\begin{table}
    \caption{Selected genera in the gut microbiome data and their corresponding covariates.}
    \begin{center}
    \begin{tabular}{llllr}
    \Hline
    &&&Selected& Averaged\\
    Phylum & Family & Genus & covariates & coefficient\\
    \hline
    Bacteroidetes & \textsl{Bacteroidaceae} & \textsl{Bacteroides} & tfat, aofib & 0.88 (0.30)\\
    Firmicutes    & \textsl{Erysipelotrichaceae} & \textsl{Catenibacterium} & tfat, aofib & 0.59 (0.15)\\
    Firmicutes    & \textsl{Family XIII} & \textsl{AD3011 group} & sex, aofib & 0.84 (0.16)\\
    Firmicutes    & \textsl{Lachnospiraceae} & \textsl{Anaerostipes} & aofib &0.82 (0.17)\\
    Firmicutes    & \textsl{Lachnospiraceae} & \textsl{Lachnoclostridium} & - & -0.88 (0.05)\\
    Firmicutes    & \textsl{Lachnospiraceae} & \textsl{NK4A136 group} & - & -0.62 (0.03)\\
    Firmicutes    & \textsl{Lachnospiraceae} & \textsl{UCG-004} & aofib & 0.58 (0.11)\\
    Firmicutes    & \textsl{Ruminococcaceae} & \textsl{Ruminococcus 2} & - & 0.85 (0.09)\\
    Firmicutes    & \textsl{Ruminococcaceae} & \textsl{UCG-002} & sex, aofib & -1.03 (0.14)\\
    Firmicutes    & \textsl{Ruminococcaceae} & \textsl{Unclassified} & sex, tfat & -0.45 (0.11)\\
    Firmicutes    & \textsl{Veillonellaceae} & \textsl{Megasphaera} & sex, tfat, aofib & 0.94 (0.14)\\
    \Hline
    \end{tabular}
    \end{center}
    \label{realdata_sel}
\end{table}

\begin{figure}
\centering
 \includegraphics[width=\linewidth, trim = {2cm 0 0 0}, clip]{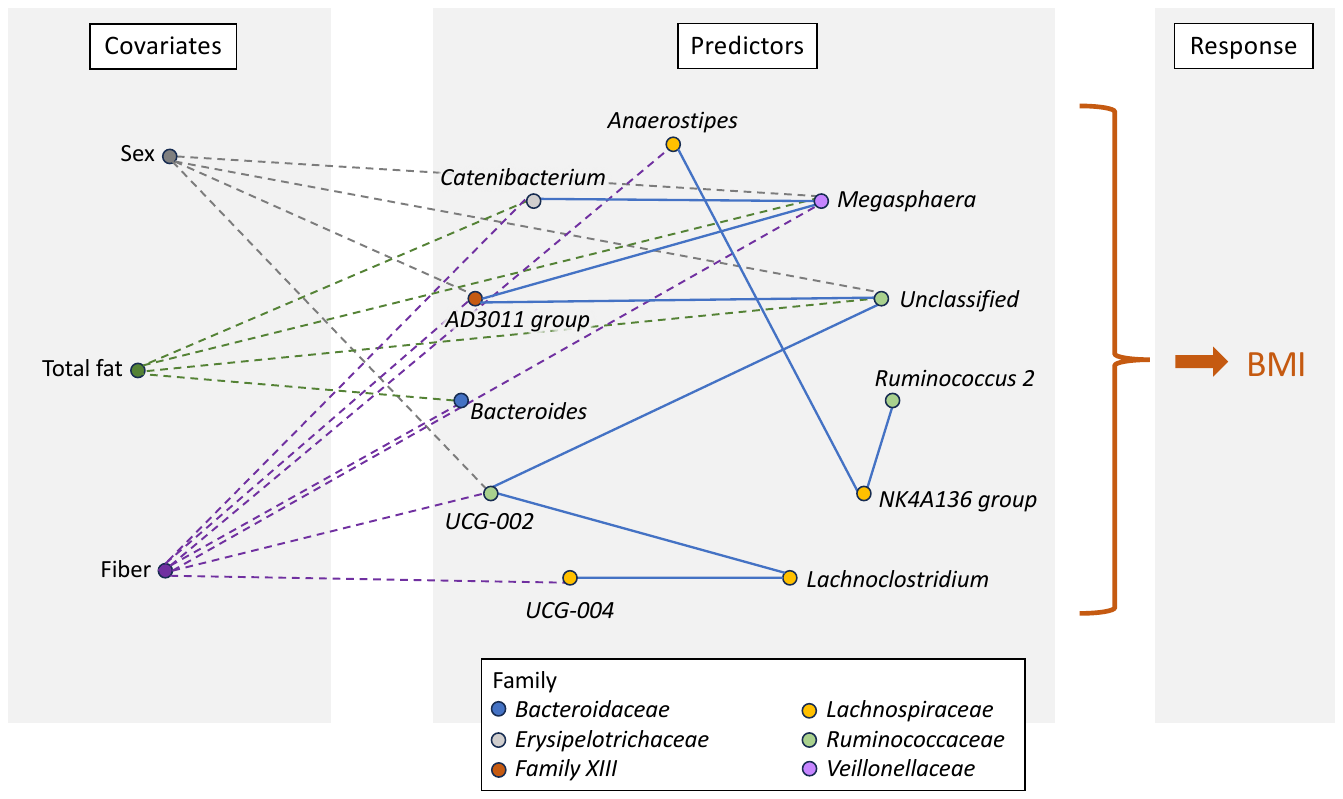}
\caption{Connections among selected predictors (genus-level microbiome features) and their corresponding covariates.} 
\label{Combo_network}
\end{figure}

In addition to identifying microbiome features, we also uncover their interrelationships and the covariates that modulate their effects. Figure \ref{Combo_network} shows the connections among selected predictors and covariates for each genus. The inferred microbiome network consists of 108 edges, including eight among the selected predictors. Several of the connections link closely related genera, consistent with previous findings that microbiome networks often show an assortative structure, where taxa close in the taxonomic tree are commonly linked in co-occurrence networks \citep{Ha2020}.

\begin{figure}
\centerline{
\includegraphics[width=15cm]{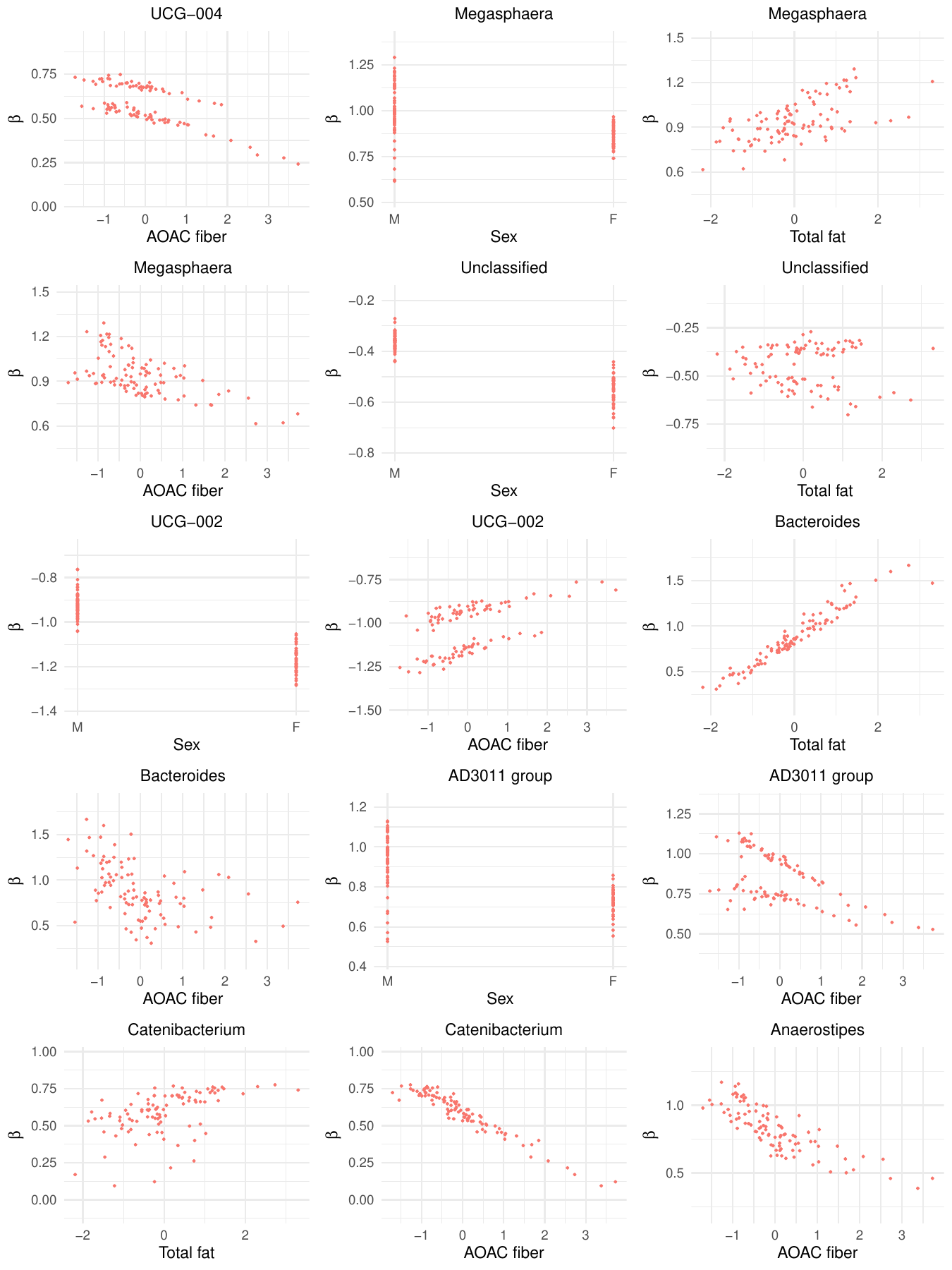}}
\caption{Estimated coefficients for selected gut microbiome and their corresponding covariates.} 
\label{Combo_cov}
\end{figure}

The effects of the selected covariates on the estimated predictor coefficients are illustrated in Figure \ref{Combo_cov}, highlighting several interesting relationships. First, increased dietary fiber intake generally attenuates the impact of microbiome features on BMI, underscoring fiber's protective role \citep{Den2013}. Second, coefficient estimates for male versus female subjects show notable separation in many plots, both directly illustrating sex effects and indirectly through a striated pattern. These variations align with studies indicating sex-dependent relationships between microbiome composition and body fat \citep{Min2019}, including sex-specific effects of \textsl{Ruminococcaceae}.

We report on comparisons to alternative methods in Web Appendix D.

\section{Conclusion}
\label{sec:conclusion}
In this study, we proposed a novel regression framework, Bayesian Varying Effects Regression with Graph Estimation (VERGE), that enables the selection of both network-linked predictor variables and covariates that modify the predictor effects, while learning network connections among these predictors. Our approach employs a Gaussian process prior, allowing for a flexible and varying effect of predictors on the outcome variable, dependent on specific covariates, and spike-and-slab priors to achieve sparsity at both the predictor and covariate levels. Simulated data demonstrate that our method can match or exceed existing methods in terms of predictor and covariate selection and prediction accuracy, particularly in identifying predictors with subtle constant effects through the use of graph information.

We applied our method to identify microbiome effects on obesity, influenced by sex and dietary factors. The selected genera and network relationships align with previously reported associations. Our analysis reveals additional nuanced relationships, particularly highlighting dietary fiber's protective role on health and how microbiome features impact BMI differently based on sex. These associations suggest correlations rather than causality due to the biological system's complexity. To confirm causal links, prospective randomized studies of microbiome and dietary interventions are necessary \citep{Durack2019}.

While our current framework focuses on continuous responses, VERGE is adaptable to other types such as binary responses or survival outcomes. Additionally, it could be expanded into a longitudinal framework by incorporating time as a varying effect. Furthermore, while the current approach uses the same set of covariates for all predictors, it could be modified to accommodate different covariates for various predictors based on prior knowledge.
 
\backmatter
\section*{Acknowledgements}
Y.R., C.B.P, and M.V. were supported in part by grants DMS-2113602 and DMS-2113557 from the National Science Foundation (NSF).

\bibliographystyle{biom} 
\bibliography{ref.bib}

\label{lastpage}

\end{document}